\documentclass[12pt]{article} 
\usepackage{amssymb} 
\usepackage{graphicx} 
\usepackage{psfig} 
\usepackage{latexsym}
\usepackage{multirow} 
\newcommand{\eq}{\begin{equation}} 
\newcommand{\en}{\end{equation}} 
\newcommand{\qe}{\end{equation}} 
\newcommand{\ear}{\begin{eqnarray}} 
\newcommand{\eqa}{\begin{eqnarray}} 
\newcommand{\rae}{\end{eqnarray}} 
\newcommand{\ena}{\end{eqnarray}} 
\newcommand{\beq}{\begin{equation}}  
\newcommand{\eeq}{\end{equation}} 
\newcommand{\bea}{\begin{eqnarray}} 
\newcommand{\eea}{\end{eqnarray}}

\newcommand{\EQ}{\begin{equation}} 
\newcommand{\EN}{\end{equation}}

\newcommand{\var}{\varepsilon}

\newcommand{\br}{\langle} 
\newcommand{\kt}{\rangle}

\newcommand{\PR}[1]{Phys.\ Rev.\ {\bf #1}}

\begin{document} 
\begin{titlepage} 
\vskip0.5cm 
\begin{flushright} 
DFTT 18/04\\ 
SISSA 59/2004/FM\\ 
IFUM-793-FT\\ 
\end{flushright} 
\vskip0.5cm 
\begin{center}  
{\Large\bf Amplitude ratios for the mass spectrum of the 2d Ising
model in the high-$T,~H \neq 0$ phase} 
\end{center} 
\vskip1.3cm 
\centerline{
M. Caselle$^a$, P. Grinza$^{b}$ and A. Rago$^c$} 
 \vskip1.0cm 
 \centerline{\sl  $^a$ Dipartimento di Fisica
 Teorica dell'Universit\`a di Torino and I.N.F.N.,} 
 \centerline{\sl via P.Giuria 1, I-10125 Torino, Italy} 
 \vskip 0.2cm 
 \centerline{\sl $^b$ SISSA and I.N.F.N, via Beirut 2-4, I-34014 Trieste, 
 Italy}  
 \vskip0.2 cm 
 \centerline{\sl  $^c$ Dipartimento di Fisica
 Teorica dell'Universit\`a di Milano and I.N.F.N.,} 
 \centerline{\sl Via Celoria 16, I-20133 Milano, Italy} 
 \vskip1.0cm 
 
\begin{abstract} 
We study the behaviour of the 2d Ising model in the symmetric 
high temperature phase in presence of a small  
magnetic perturbation. We successfully compare the quantum  
field theory  predictions for the shift in the mass spectrum of the theory   
with a set of high precision transfer matrix results. Our results rule out 
a prediction for the same quantity obtained some years ago
with strong coupling methods.    
\end{abstract} 
\end{titlepage} 
 
\setcounter{footnote}{0} 
\def\thefootnote{\arabic{footnote}} 
 
\section{Introduction.} 
\label{sect1} 
Despite its apparent simplicity and the fact that since 1944  
the exact expression of the free energy along the $h=0$ axis is known 
exactly~\cite{o44} the two dimensional Ising model is still  
an endless source of interesting and challenging 
problems~\cite{z89}-\cite{gr_abo}. 
 
In particular in these last few years a renewed interest has been attracted by 
the study of the model in its whole complexity, i.e. in presence of both magnetic 
and thermal perturbations. There are two main reasons behind this  renewed 
interest. On one side recent progress in 2d quantum field theory (in particular, 
but not only, in the framework of perturbed CFT's and S-matrix integrable models) 
allowed to obtain a host of new important results on correlation functions, 
amplitude ratios and more generally on perturbations around integrable 
directions~\cite{z89}-\cite{cggm03}. On the other side the appearance of new powerful  algorithms (both 
for montecarlo simulations and for transfer matrix studies) and the increasing 
performances of computers allowed to precisely test a lot of the above 
predictions~\cite{gr03}-\cite{gr_abo}.  
This paper, which is devoted to the study of Ising model in presence of a small 
magnetic perturbation around the integrable thermal axis, 
is a further step in this direction. It can be considered as the natural 
continuation of~\cite{gr03} in which, with similar tools,  
the opposite setting of a small thermal perturbation around the magnetic 
integrable line was studied and it is part of our more general ongoing 
project~\cite{gm03}-\cite{gr_abo} 
on the study and characterization  
of the 2d  Ising model in the whole thermal and magnetic plane. A further reason 
of interest for the present 
 analysis is the theoretical estimate recently obtained by 
Fonseca and Zamolodchikov~\cite{fz03} 
 for the first order magnetic correction to the mass 
spectrum of the model. This estimate turns out to be in sharp disagreement
(more than one order of magnitude) with
a previous estimate of the same constant obtained more than 20 years ago with
strong coupling methods in~\cite{sss81}. This disagreement is very
puzzling, since  strong coupling estimates in the 2d
Ising model are  usually highly reliable.  
 One of the goals of the present work is to fix this disagreement testing these
 two estimates with a high precision transfer matrix calculation directly in the
 2d Ising model. 
This paper is organized as follows. In sect.~\ref{sect2} we first  
recall some known results on the 2d Ising model and then concentrate on the 
definitions and theoretical predictions for the 
 quantities which we measure in our transfer matrix analysis. Sect.~\ref{sect3} 
 is devoted to a short discussion of the transfer matrix method while in the 
 last section we discuss our results and compare our findings with the 
 theoretical predictions.    
 
\section{The model.} 
\label{sect2} 
In this section we briefly recall some well known results on the Ising model. 
A more detailed description can be found in the 
reviews\cite{mw73}-\cite{d04}\footnote{The book~\cite{mw73} is the standard 
reference for the lattice Ising model. An updated version by one of the authors 
 can be found in~\cite{mc95}. A recent thorough review of the field theoretic 
 approach to the model can be found in~\cite{d04}.}. 
 
\subsection{The lattice version of the model.} 
\label{sect2.1} 
The  Ising model in a magnetic field is defined 
by the partition function 
\eq 
Z=\sum_{\sigma_i=\pm1}e^{\beta(\sum_{\br n,m \kt}\sigma_n\sigma_m 
+H\sum_n\sigma_n)} 
\label{zz1} 
\en 
where the field variable $\sigma_n$ takes the values $\{\pm 1\}$; 
$n\equiv(n_0,n_1)$ labels the sites of a square lattice  size $L_0$ and $L_1$ 
in the two directions and lattice spacing $a$\footnote{Since the lattice spacing 
will play no role in the following we shall set $a=1$ in the rest of the  
paper.}. 
 $\br n,m \kt$  
denotes nearest neighbor sites on the lattice. 
In the following  
we shall treat asymmetrically the two directions. We shall denote 
$n_0$ as the compactified  ``time'' coordinate and $n_1$ as the space one. 
The number of sites  of 
the lattice will be denoted by  $N\equiv L_0 L_1$. The lattice extent in the 
transverse (``time") direction will be denoted as $L_0$.  
Following the standard notation we  define  $h_l=\beta H$ thus the partition 
function becomes:  
\eq 
Z(h_l)=\sum_{\sigma_i=\pm1}e^{\beta\sum_{\br n,m \kt }\sigma_m\sigma_m 
+h_l\sum_n\sigma_n}   \;\;\; . 
\label{fupart} 
\en 
 
As it is well known the 2d Ising model has a second order phase transition at 
$h_l=0$ and $\beta=\beta_c$ given by: 
 $$\beta=\beta_c=\frac12\log{(\sqrt{2}+1)}=0.4406868...$$ 
 
In the following we shall be interested in the high temperature phase of the 
model (i.e. $\beta<\beta_c$) in which the $\mathbb{Z}_2$ symmetry is unbroken. 
 
\subsection{Continuum theory} 
\label{sect2.2} 
In the continuum limit the model is described 
by the action: 
\EQ 
{\cal A}={\cal A}_{0}-\tau\int d^2x\,\var(x)-h\int d^2x\,\sigma(x)\,, 
\label{A} 
\EN 
where $\sigma(x)$ and $\var(x)$ denote the magnetic and thermal  
perturbing operators respectively while $h$ and $\tau$  
are dimensional couplings measuring the magnetic field and  
the deviation from critical temperature. 
A standard scaling analysis shows that the scaling behaviour of these two 
couplings is 
\bea 
&& \tau\sim m^{2-x_\var}=m\,, \nonumber\\ 
&& h\sim m^{2-x_\sigma}=m^{15/8} 
\label{scaling} 
\eea 
where $m\sim 1/\xi$ ($\xi$ being the correlation length of the model) 
denotes the mass  
scale associated to the breaking of scale invariance away from criticality,
$x_\var=1$ and $x_\sigma=1/8$ are the scaling dimensions of the energy and spin
operators respectively and  
${\cal A}_0$ denotes the conformal invariant action at the critical point. 
In the Ising case, which is known to be described by a free massless Majorana 
fermion, this can be written explicitly as: 
\EQ 
{\cal A}_0=\frac12\int d^2x\,(\psi\bar{\partial}\psi+ 
\bar{\psi}\partial\bar{\psi})\,, 
\label{A0} 
\EN 
where $\partial=\partial_z=(\partial_1-i\partial_2)/2$ and  
$\bar{\partial}=\partial_{\bar{z}}=(\partial_1+i\partial_2)/2$ and  
the operators $\psi$ and $\bar{\psi}$ are 
the two components of the neutral Majorana fermion. 
 
The thermal perturbing operator (which coincides with the energy operator) 
can be written as 
\EQ 
\varepsilon\sim\bar{\psi}\psi\,; 
\label{energy} 
\EN 
which allows to identify it as the mass term of the free fermionic action, in 
agreement with the first of the scaling equations (\ref{scaling}). 
 
By combining the two scaling equations (\ref{scaling}) one can see that 
the field theory (\ref{A}) describes a one-parameter family of renormalization  
group trajectories flowing out of the critical point at $\tau=h=0$  
and labeled by the dimensionless quantity\footnote{To avoid confusion let us  
stress that we follow the definition for $\eta$ of~\cite{d04} which is different 
from the one adopted in~\cite{fz03}.} 
\EQ 
\eta=\frac{\tau}{|h|^{8/15}}\,\,. 
\label{eta} 
\EN 
 
The Ising field theory can be solved exactly in the two limiting cases of $h=0$  
and $\tau=0$. In the first case, as we have seen above, it is simply the theory 
of {\em free massive} fermion, the mass  
being proportional to $|\tau|$. In the second case (which corresponds in the 
lattice discretization to the magnetic perturbation with $\beta$ fixed to the 
critical value 
$\beta=\beta_c$)  
A.~Zamolodchikov was able to show that (\ref{A})  
with $\tau=0$ is a complicated but 
 integrable quantum field theory \cite{z89} of eight interacting particles.  
 
For generic values of $h$ and $\tau$ exact integrability is lost but 
notwithstanding this some important theoretical results can all the same  
be obtained. Let us see a few of them. 
\begin{itemize} 
\item {\bf Equation of state.} 
 
With a combination of transfer matrix techniques and analytic continuation of 
suitable parametric representations it is possible to construct a very precise 
expansion for Helmholtz free energy of the the model and from it of the equation 
of state in the whole critical region in the $(\tau,h)$ plane~\cite{chpv01}. From 
this expression precise predictions for several 
 universal amplitude ratios can be obtained (see~\cite{chpv01} for details). 
 
\item {\bf The spectrum of the model.} 
This important result was conjectured for the first time in a 
 seminal paper of Mc~Coy and Wu~\cite{mw78} and  recently re--understood  
 in a field theoretic language (see \cite{d04} for a thorough discussion). 
The scenario which emerges is the following. For $\eta=-\infty$ (i.e. in the 
low temperature broken symmetry phase)  the spectrum starts with a cut. In this 
limit the natural degrees of freedom of the model are kinks  
interpolating between the two degenerate vacua which  
are non-local with respect to the spin degrees of freedom of the model. As the 
magnetic field is switched on (i.e. for large and negative values of 
$\eta$) this cut breaks up in a host of particles which may be understood as 
bound states of the above mentioned kinks. As $\eta$ is increased from $-\infty$  
to finite negative values the number of stable particles (i.e. below the pair 
creation threshold $m_n<2m_1$)  
decreases. Exactly at $\eta=0$ only three such particles are left  
(but, only for this value of $\eta$, due to the exact integrability of the model 
five more particles turn out to be stable even if they are above threshold).  
The number of stable particles continues to decrease as $\eta$ increases, until a single particle is 
left at large enough $\eta$. There is a wide region of values of $\eta$ before the positive thermal axis 
(i.e. $\eta=+\infty$) is reached in which the spectrum contains only one stable particle. This is the region 
in which we shall perform our analysis in the following.
\begin{figure}[htb]
\centerline{\psfig{figure=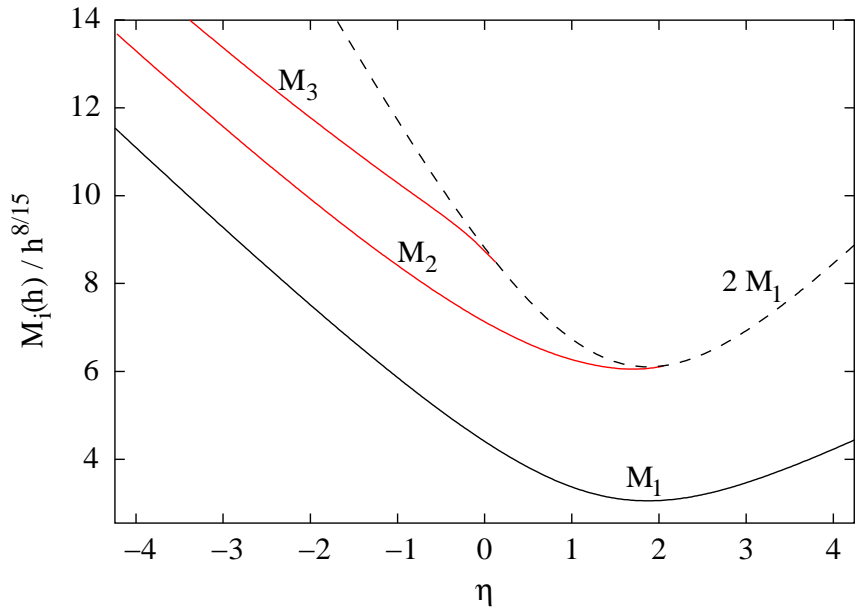,width=0.7\textwidth}}
\caption{Ising Model's masses, the value of the masses has been obtained in \cite{fz2001}}
\label{masses}
\end{figure}
\item {\bf Small perturbations around the integrable lines.} 
The most interesting result for our present purposes is that in the nearby of 
the two integrable lines a few exact results can be obtained using perturbative 
methods. In particular one can evaluate the corrections to the masses and to the 
ground state energy due to the perturbing operator~\cite{dms96}.  
Again we have two possible 
situations: 
\begin{description} 
\item{a]} A small thermal perturbation of the Zamolodchikov's integrable model 
along the $\tau=0$ magnetic axis. Estimates for the first order mass  
correction where obtained in~\cite{dms96} and  successfully tested both on the 
lattice, using transfer 
matrix methods~\cite{gr03} and directly in the Ising field theory  
by a numerical diagonalization of the 
Hamiltonian on a conformal basis of states~\cite{dms96}.  
\item{b]} 
A small magnetic perturbation of the free fermionic theory along the $h=0$ axis. 
This case was recently studied by Fonseca and Zamolodchikov~\cite{fz03} which 
were able to obtain, using the Ward identities of the model, the matrix elements 
of the product $\sigma(x)\sigma(x')$ between any particle states. This is 
exactly the ingredient which is needed to obtain the sought for mass correction. 
 The aim of our paper is to test this result with a transfer matrix analysis 
 directly on the lattice model. We shall devote the next subsection to a 
 detailed discussion of this term. 
\end{description} 
\end{itemize} 

\subsection{The mass correction} 
\label{sect2.3} 
Following the above discussion 
in the $\tau>0$ phase, for small value of $h$  there is  a single stable 
particle in the spectrum. Let us call $m(\tau,h)$ its mass. 
For symmetry reasons  
the correction to the mass of the particle when the magnetic field is  
switched on at $\tau>0$  must be proportional to $h^2$. Scaling arguments (see 
the discussion in the previous section) then fix the dependence on $\tau$ 
to be $\tau^{-15/4}$. 
Thus we expect the following behaviour: 
\eq 
m(\tau,h)=m(\tau,0)(1+\hat a h^2 \tau^{-15/4}) 
\label{eq:1} 
\en  
From a numerical point of view it is useful to rewrite this dependence as a 
function of the unperturbed mass $m(\tau,0)$: 
\eq 
m(\tau,h)=m(\tau,0)(1+ a h^2 m(\tau,0)^{-15/4}) 
\en  
Recently, in a remarkable paper Fonseca and Zamolodchikov \cite{fz03} were able to evaluate analytically  
this constant which turns out to be  
\bea 
a=10.7619899(1) 
\label{acost}
\eea
The above correction can also be rewritten in a form which will be useful in the 
following:  
\eq 
m^2(\tau,h)=m^2(\tau,0)+ \tilde a h^2 \tau^{-7/4} 
\en  
These three constants, which are actually the same constant written in different
units, can be easily related among them. In fact,
thanks to the fact that the Ising quantum field theory is solved exactly along 
the thermal line, the 
relation between $\tau$ and $m(\tau,0)$ is know exactly\footnote{This relation, 
as all the ones which follow in this section actually depends  
on the conventions which one chooses for the structure constants.  
In this paper we follow the so called ``conformal normalization'' i.e.  
$C_{\sigma\sigma}^I=C_{\varepsilon\varepsilon}^I=1$~.}: $m=2\pi \tau$. 
This allows to write the following relations:
\begin{eqnarray}
\tilde a&=&2a (2\pi)^{-7/4}\\
\hat a&=&a (2\pi)^{-15/4}
\end{eqnarray}
\subsubsection{Theoretical estimates for  mass correction} 
Up to a few months ago the only existing estimate for the constant $a$ was
the one reported in~\cite{sss81}, obtained  performing the continuum limit 
extrapolation of the 
strong coupling expansion of the perturbed two point correlator~\footnote{
We have normalized the result of~\cite{sss81} in units of the continuum field theory discussed above, thus it may be directly compared
with the one of eq.~(\ref{acost}).}:
\eq
a=(\frac14+\frac{1}{\pi})(2\sqrt{2})^{\frac14}= 0.73700676...
\en
This estimate
was never tested with numerical simulations or transfer matrix methods.
Further, 
the impressive discrepancy between (\ref{acost}) and the above estimate was one of the major 
reasons which prompted us to perform the present
transfer matrix analysis.

\subsection{Correction in the free energy.} 
Similarly to the mass case
 one can define the deviation from the unperturbed value of the free 
energy ${\cal F}$ (which is nothing else than the lowest eigenvalue of the transfer
matrix spectrum). In this case it is easy to obtain a theoretical estimate of
this deviation by noticing that for symmetry reasons
 the first nonzero 
correction must be again at order $h^2$ and is exactly given by $1/2$ times  
the magnetic susceptibility 
\eq 
{\cal F}(\tau,h)={\cal F}(\tau,0)+\frac{h^2}{2}\chi_+(\tau,0)
+{\cal O}(h^4)\equiv {\cal F}(\tau,0)+ \frac{\Gamma_+}{2}
 h^2 \tau^{-7/4}+... 
\label{eq:2}\en 
Where in the last term of the above equation  $\Gamma_+$ denotes the 
susceptibility amplitude and
 we have inserted the known dependence on $\tau$ of the susceptibility. 
An important role in the following analysis is played by the fact that 
the value of $\Gamma_+$ in the the continuum theory is known 
exactly~(see for instance~\cite{d04}):
\eq 
\Gamma_+ =0.148001214... \, .
\en 
 
\subsection{Universal amplitude ratios.} 
A major problem in comparing the above values, obtained in the continuum limit
QFT,
with our numerical estimates in the 
lattice model is that we must convert the  
continuum limit parameters $\tau$ and $h$ into their lattice counterparts $t$ 
and $h_l$.  
A nice way to avoid this problem is to combine the above constants in a suitable 
universal ratio which is thus independent from the particular realization of the 
underlying quantum field theory. There are two natural choices:
\begin{itemize}
\item
Following~\cite{d04} we shall study the
combination  
\eq 
\label{ratio}
R_1=\lim_{\eta\to\infty}\frac{\delta m^2}{\delta{\cal F}} 
\en 
where 
\eq 
\delta m^2\equiv m^2(\tau,h)-m^2(\tau,0) 
\label{defdm2}
\en 
\eq 
\delta {\cal F}\equiv {\cal F}(\tau,h)-{\cal F}(\tau,0) 
\label{deff}
\en 
A direct substitution gives: 
\EQ 
\lim_{\eta\to\infty}\frac{\delta m_1^2}{\delta{\cal F}}=\frac{4 a}
{\Gamma_+(2\pi)^{7/2}}
\EN 
\item
A second option is:
  \eq
 R_2=\frac{\delta m}{m(\tau,0)}\frac{M_s^2(-\tau)}{M^2(\tau,h)}
  \en
where 
\eq 
\delta m\equiv m(\tau,h)-m(\tau,0) 
\label{defdm}
\en 
while $M(\tau,h)$ denotes the magnetization and $M_s(-\tau)$ the spontaneous
magnetization.
A direct substitution of the various terms leads to
\eq
R_2=a(2\pi)^{-15/4}\frac{B^2}{\Gamma^2_+}
\en
where $B$ denotes the amplitude of the spontaneous magnetization whose value is
 known exactly  in the continuum limit theory (see for instance~\cite{d04})
\eq
B=1.70852190.
\en
\end{itemize}
The expected values for these ratios, using the Fonseca-Zamolodchikov estimate
for the constant $a$ are
\bea
R_1=11.66467...\hskip2cm
R_2=1.4568962...
\label{FZratio}
\eea
while with the strong coupling estimate for $a$ we should expect:
\eq
R_1=0.7988246...\hskip2cm
R_2=0.09977173...
\en
\subsection{Lattice amplitudes.} 
\label{2.6} 
This is all we can do for a generic realization of the Ising quantum field 
theory. However in the particular case of the 2d square lattice realization 
thanks to the fact that the model is exactly solved {\sl also on the lattice} 
for $h=0$ we know the value  of several 
of the above amplitudes directly in the lattice realization.
The mass of the theory in the transfer matrix geometry is
\eq
m_\ell=-\log~v\left(\frac{1+v}{1-v}\right)
\label{eq:ml}
\en
with $v=\tanh \beta$. The index $\ell$ recalls that this is the lattice estimate of
the mass of the model. Eq.(\ref{eq:ml}) 
leads to the following expansion
 in the vicinity of the critical point
\eq
m_\ell(t,0)\sim 4\beta_c t
\en

where $t$ is the reduced temperature defined as
\eq
t=\frac{|\beta-\beta_c|}{\beta_c}.
\en

The spontaneous magnetization is given by
\eq
M_s=\left(1-\frac{1}{sh^4(2\beta)}\right)^{\frac18}
\en
which gives:
\eq
M_s\sim(-8\sqrt2 \beta_c~t)^{\frac18}.
\en

For 
the magnetic susceptibility $\chi_+ (t)$ there is no exact expression. However this
observable has been studied extensively in the past years, firstly by McCoy et al. 
\cite{Wu:1975mw} and more recently by Orrick et al. \cite{ongp} using strong coupling expansions (notice
that the
first  term in the expansion in powers of $t$ can be evaluated exactly). One finds:
\bea
\chi_+ (t) = C_{0,+} t^{-7/4} + C_{1,+} t^{-3/4} +\dots 
\eea
where
\bea
C_{0,+} & = & 0.962581732 \dots \nonumber \\ 
C_{1,+} & = & 0.074988153 \dots \; . 
\label{numes}
\eea
Plugging these amplitudes (together with the lattice estimate of the constant $a$ which we shall discuss in
the following section) it is possible to obtain the two ratios $R_1$ and $R_2$ discussed in the previous
section. 

Another possible use of these results 
is to construct the explicit relation between $t$ and $\tau$ and 
similarly between $h_\ell$ and $h$. This will allow us to test separately the two 
predictions  eq.(\ref{eq:1}) and (\ref{eq:2}). Comparing the lattice and continuum values for the mass we
find:
\eq
\tau  =  \frac{\log(1+\sqrt 2)}{ \pi} t \equiv k_\tau \; t, \ \ \ \
\en
while comparing the susceptibility amplitudes 
\eq
C_{0,+} =  k_\tau^{-7/4} k_h^2 \; \Gamma_+ 
\en
we find
\eq
h   =  2^{5/48} e^{1/8} A^{-3/2} h_l  \equiv k_h \; h_\ell
\en 
(where $A=1.28242712 \dots$ is the Glaisher constant).

With these results at hand we may write an explicit prediction for the constant $a$ directly on the lattice.
Defining the lattice version of $a$ as:
\eq
a_\ell\equiv k_h^2 k_\tau^{-15/4}~a
\en
we find using the value for $a$ obtained by Fonseca and Zamolodchikov~\cite{fz03}
\bea
a_\ell=7.56977 \dots
\label{FZesti}
\eea
and, using the strong coupling one discussed in~\cite{sss81}:
\bea
a_\ell=0.518396 \dots \, .
\label{turcofollesti}
\eea
In the following we shall use this last result and shall directly extract from the transfer matrix data the value of $a_\ell$ which we shall
then combine with the other amplitudes in order to obtain the two universal ratios $R_1$ and $R_2$.

\section{The transfer matrix analysis} 
\label{sect3} 
For the analysis of the transfer matrix data we  followed the same procedure of our previous works \cite{gr03,ch99,gr_abo} which can be schematically sketched as follow.
\begin{itemize}
\item We evaluate the first few eigenvalues of the transfer matrix for the values of temperature and magnetic coupling in which we are interested for lattice widths in the transverse direction in the range  $L\in\{9-21\}$.
We construct from them the observables in which we are intested (mass and free energy) and extrapolate their infinite lattice limits using the recursive procedure discussed in detail
 the appendix of \cite{Caselle:2000bj}.
\item For each values of $t$ and $h_\ell$ 
we construct the combinations $\delta m$ and $\delta m^2$ and $ 
\delta {\cal F}$
defined in eq.(\ref{defdm}), (\ref{defdm2}) and (\ref{deff}):  
\item
We fit $\delta m^2$, $\delta m$ and $\delta {\cal F}$ 
assuming the known scaling behaviour for these quantities. As we shall see the first few
terms of the scaling functions will be enough to obtain stable results for the fits within the precision of our data.
\end{itemize}
In table (\ref{par_val}) we report the parameters and the various settings we used in our simulations, while in tables 
(\ref{extrapol},\ref{fitting}) we report a sample of our data to allow the interested reader to reproduce our analysis.
\begin{table}[th]
\begin{center}
\begin{tabular}{lll}
\hrulefill&\hrulefill&\hrulefill\\[-4.5mm]
\hrulefill&\hrulefill&\hrulefill\\[-1.5mm]
\multicolumn{1}{c}{$t$}& \multicolumn{1}{c}{$h_\ell$}&  \multicolumn{1}{c}{$L$}\\[-4mm]  
\hrulefill&\hrulefill&\hrulefill\\[-4.5mm]
\hrulefill&\hrulefill&\hrulefill\\[-1mm]
0.345&0.0&9\\
0.350&0.0003&10\\
0.355&0.0005&11\\
0.360&0.0008&\multirow{3}{5mm}{$\vdots$}\\
0.365&0.0010&\\
0.370&0.0013&\\
&0.0015&19\\
&0.0018&20\\
&0.0020&21\\
\end{tabular}
\end{center}
\caption{Parameters value.}
\label{par_val}
\end{table}
\begin{table}[bth]
\begin{center}
\protect\footnotesize
\begin{tabular}{|l|l|l|l|}
\hline
$L$&$1/m(t,h_\ell)$&${\mathcal F(t,h_\ell)}$&$M_\ell(t,h_\ell)$\\
\hline\hline
15&2.565917754587&0.830031898983&0.020167928578\\
16&2.565918168084&0.830031896272&0.020167777385\\
17&2.5659182662035&0.830031895299&0.020167731420\\
18&2.565918247605&0.830031894746&0.020167716025\\
19&2.565918227627&0.830031894142&0.020167710600\\
20&2.565918215050&0.830031894612&0.020167708951\\
21&2.565918209646&0.830031893614&0.020167708217\\
\hline\hline
$\infty$&2.56591821(1)&0.830031894(1)&0.020167708(1)\\
\hline
\end{tabular}
\end{center}
\caption{Example of data used for the extrapolation ($h_\ell=0.0013$, $\beta=0.350$).}
\label{extrapol}
\end{table}
\section{Discussion of our results.} 
\label{sect4} 
Now we are in the position to compare the lattice results, coming from the transfer matrix approach, with the field theoretic estimates of the ratios $R_1$, $R_2$. To this aim, the first step is to 
extract the leading behaviour of $\delta m$, $\delta m^2$, $\delta {\mathcal F}$ and $M_\ell(t,h_\ell)$ 
from the data obtained by means of the transfer matrix.

As a preliminary check of our procedure we estimated from the transfer matrix data the mass at zero magnetic field
which in this geometry
is known exactly and is given by eq.(\ref{eq:ml}). We report in tab.\ref{tabnew} the comparison which turns out to be fully satisfactory.

Then we concentrated in the evaluation of $\delta m^2$. Symmetry considerations suggest 
 the following expansion in even powers of $h_\ell$ for this quantity.
\bea
\delta m^2 = f_1 (t) h_\ell^2 +  f_2 (t) h_\ell^4 + \dots
\eea    
It turns out that, within the precision of our data and given our choice of values of $h_\ell$, 
it is enough to take into account only the first two terms in this fit, the term
proportional to $h^6$ being completely negligible.
For each fixed value of $t$ we extract from the fit the  best fit values for $f_1(t)$ and $f_2(t)$.  
Since we are interested in the leading behaviour of $\delta m^2$, 
in the following we shall concentrate in the study of 
the function $f_1 (t)$. The  functional form of $f_1(t)$ can be easily constructed using
the same RG and CFT arguments 
developed in \cite{Caselle:2001jv}-\cite{Caselle:2000bj}, \cite{gr03}:
\bea
f_1 (t) = \tilde{a}_{\ell} t^{-7/4} + a_1 t^{-3/4} + a_2 t^{1/4} + \dots
\eea 
where $\tilde{a}_{\ell} $ is the amplitude in which we are interested.\\
Following the same fitting procedure used in \cite{gr03}, we obtain:
\bea
\tilde{a}_{\ell} = 5.61 \pm 0.01
\eea
which gives
\bea
a_{\ell} = 7.56 \pm 0.01 
\eea
in complete agreement with the Fonseca-Zamolodchikov estimate of the amplitude $a$ 
(see eq.~(\ref{FZesti})). 
\begin{table}[!t]
\begin{center}
\protect\footnotesize
\begin{tabular}{|l|l|l|l|l|}
\hline
$h_\ell$ &$\beta=0.345$ &  $\beta=0.350$ &$\beta=0.355$ &$\beta=0.360$  \\
\hline \hline 
0&2.42236580(1)&2.56721580(1)&2.72886532(1)&2.91043999(6)\\
0.0003&2.42231285(1)&2.56714664(1)&2.72877368(1)&2.91031661(4)\\
0.0005&2.42221873(1)&2.56702370(1)&2.72861080(1)&2.91009730(5)\\
0.0008&2.42198936(1)&2.56672412(1)&2.72821391(1)&2.90956300(5)\\
0.0010&2.42177770(1)&2.56644770(1)&2.72784773(1)&2.90907012(4)\\
0.0013&2.42137222(1.5)&2.56591821(1)&2.72714641(1)&2.90812629(4)\\
0.0015&2.42104331(1)&2.56548878(1)&2.72657770(2)&2.90736112(3)\\
0.0018&2.42046226(1)&2.56473025(1)&2.72557340(1)&2.90601025(2)\\
0.0020&2.42001655(1)&2.56414851(1)&2.72480337(1)&2.90497486(3)\\
\hline
\multicolumn{5}{c}{}\\[-3.5mm]
\multicolumn{5}{c}{(a) Inverse mass $1/m(t,h_\ell)$}\\
\end{tabular}\\
\vspace{3mm}
\begin{tabular}{|l|l|l|l|l|}
\hline
$h_\ell$ & $\beta=0.345$ &$\beta=0.350$ & $\beta=0.355$& $\beta=0.360$\\
\hline \hline 
0.&0.825668076(1)&0.830018782(1)&0.8344669273(3)&0.839014965(1)\\
0.0003&0.825668712(1)&0.830019481(1)&0.834467698(1)&0.839015821(1)\\
0.0005&0.825669843(1)&0.830020722(1)&0.834469068(1)&0.839017342(1)\\
0.0008&0.825672600(1)&0.830023748(1)&0.834472408(1)&0.839021049(1)\\
0.0010&0.825675145(1)&0.830026541(1)&0.834475490(1)&0.839024471(1)\\
0.0013&0.825680021(1)&0.830031894(1)&0.834481397(1)&0.839031028(1)\\
0.0015&0.825683978(1)&0.830036238(1)&0.834486189(1)&0.839036348(1)\\
0.0018&0.825690973(1)&0.830043915(1)&0.834494660(1)&0.839045750(1)\\
0.0020&0.825696341(1)&0.830049807(1)&0.834501162(1)&0.839052965(1)\\
\hline
\multicolumn{5}{c}{}\\[-3.5mm]
\multicolumn{5}{c}{(b) Free energy ${\mathcal F} (t,h_\ell)$}\\
\end{tabular}\\
\vspace{3mm}
\begin{tabular}{|l|l|l|l|l|}
\hline
$h_\ell$ & $\beta=0.345$ &  $\beta=0.350$ & $\beta=0.355$&$\beta=0.360$\\
\hline\hline
0.0003&0.004241449(1)&0.004655977(1)&0.005138273(1)&0.00570442(1)\\
0.0005&0.007068848(1)&0.00775956(1)&0.00856335(1)&0.00950676(1)\\
0.0008&0.011309247(1)&0.012414204(1)&0.01369968(1)&0.01520845(1)\\
0.0010&0.014135509(1)&0.015516337(1)&0.017122651(1)&0.019007841(5)\\
0.0013&0.018373549(1)&0.020167708(1)&0.022254596(1)&0.024703401(2)\\
0.0015&0.021197803(1)&0.023267128(1)&0.0256738415(5)&0.028497574(2)\\
0.0018&0.025432177(1)&0.027913548(1)&0.030798990(1)&0.034183633(2)\\
0.0020&0.028253554(1)&0.031009083(1)&0.034212903(1)&0.037970351(1)\\
\hline
\multicolumn{5}{c}{}\\[-3.5mm]
\multicolumn{5}{c}{(c) Magnetization $M(t,h_\ell)$}\\
\end{tabular}
\caption{Extrapolated values of different operators as function of the magnetic coupling and the temperature.}
\label{fitting}
\end{center}
\vspace{-5mm}
\end{table}

\begin{table}
\begin{center}
\begin{tabular}{|c|l|l|}
\hline
$\beta$ & Exact mass formula & Transfer matrix \\
\hline \hline
$0.345$ & $2.4223658021832$ & $2.42236580(1)$ \\
$0.350$ & $2.5672157958004$ & $2.56721580(1)$ \\
$0.355$ & $2.7288653104825$ & $2.72886532(1)$ \\
$0.360$ & $2.9104399518676$ & $2.91043999(5)$ \\
$0.365$ & $3.1158911582069$ & $3.115891(1)$ \\
$0.370$ & $3.3502882801313$ & $3.350287(1)$ \\
\hline
\end{tabular} 
\caption{Comparison of our TM extrapolation with the exact result for the invers mass at zero magnetic field}
\label{tabnew}
\end{center}
\end{table}
\ 

The amplitude of the leading behaviour of $\delta m$ can be immediately inferred 
from that of $\delta m^2$
\bea
\delta m   &  = &  \frac{\tilde{a}_{\ell} }{2 m_\ell(t,0)} t^{-7/4} h_\ell^2 + \dots  \\
& = & \frac{\tilde{a}_{\ell}}{8 \beta_c} t^{-11/4} h_\ell^2 + \dots  \nonumber \\
& \equiv & \tilde{b}_{\ell} \, t^{-11/4} h_\ell^2 + \dots 
\eea   
plugging in the last definition the above obtained value of $\tilde a_\ell$ 
we obtain
\bea
\tilde{b}_{\ell} = 1.591 \pm 0.002 . 
\eea
As a consistency check, we computed the same amplitude directly from the fit of the transfer matrix 
data. The final result is 
\bea
\tilde{b}_{\ell} = 1.592 \pm 0.003 
\eea
which is completely equivalent to the previous estimate.

To estimate  the vacuum energy on the lattice let us notice first of all that $\delta {\mathcal F}$
can be expanded as 
\bea
\delta {\mathcal F}  = \frac{1}{2} \chi_+ (t) h_\ell^2 + g_2(t) h_\ell^4 + \dots
\eea
where the function $\chi_+ (t)$ is the magnetic susceptibility of the model in the high-temperature phase $\beta<\beta_c$ and $h=0$ already discussed in section \ref{2.6}. 
Hence the exact estimate of the amplitude $\tilde c_\ell$ of the free energy on the lattice is given by
\bea
\tilde{c}_\ell = \frac{C_{0,+}}{2} =0.48129086 \dots \;  . 
\eea
We also compared the previous exact results (quoted in sect.~\ref{2.6}) with our numerical estimates. 
We found
\bea
C_{0,+}^{\textrm{\tiny num}} & = & 0.962 \pm 0.002 \nonumber \\
C_{1,+}^{\textrm{\tiny num}} & = & 0.075 \pm  0.001 \nonumber
\eea
in complete agreement with (\ref{numes}). 

Finally, we need the amplitude $\tilde d_\ell$ of the scaling behaviour of the magnetization.
It can be easily computed (exactly) from the definition of $M_\ell(t, h_\ell)$
\bea
M_\ell(t, h_\ell) = \frac{\partial \mathcal{F}}{\partial  h_\ell} = C_{0,+} \, t^{-7/4} h +
\dots
\eea 
leading to 
\bea
\tilde d_\ell = C_{0,+} =  0.962581732 \dots \, .
\eea
Since the transfer matrix approach allows also the numerical computation of the magnetization, we can extract the numerical estimate of $\tilde d_\ell$
\bea
\tilde d_\ell = 0.961 \pm 0.004
\eea
which, once again, agrees with the previous results. 
We also stress that the zero-field quantities $m_\ell$ and $M_s$ involved in the ratio $R_2$ are given by their exact expressions quoted in section \ref{2.6}. 

The final step is then to compute the lattice estimates of the ratios $R_1$, $R_2$
\bea
R_1 & =& \frac{\tilde{a}_\ell}{\tilde{c}_\ell} = 11.66 \pm 0.02
\nonumber \\
R_2 & =& 2^{-9/8}\beta_c^{-3/4} \frac{\tilde{b}_\ell}{\tilde{d}_\ell^2} 
= 1.456 \pm 0.002
\eea      
which agree with the field theoretic predictions obtained using the Fonseca and Zamolodchikov 
estimate of the constant $a$ (see eq.~(\ref{FZratio})).

The precision of our numerical 
computations allows to rule out unambiguosly the strong coupling estimates of these universal ratios. 

It would be interesting to investigate the reasons of this failure, which is even more surprising given the usual high reliability of
strong coupling results in two dimensional statistical models (and in particular in the Ising model).
A possible explanation is that the result of \cite{sss81} is based on a peculiar mutipole
 expansion of the perturbed two point function, which was
proved to have  good converegence properties when tested in the case of the $H=0$ susceptibility (see \cite{Wu:1975mw}) but most probably has a
less good behaviour if one is interested in the mass shift.   

\vskip1.0cm 
{\bf  Acknowledgments.} 
We would like to thank G. Delfino for useful discussions. 
This work was partially supported by the 
European Commission TMR programme HPRN-CT-2002-00325 (EUCLID). The work of P.G. is supported by the COFIN ``Teoria dei Campi, Meccanica Statistica e Sistemi Elettronici''.

\end{document}